\documentclass[12pt]{article}
\usepackage{amssymb}

\def\on#1#2{\mathop{\vbox{\ialign{##\crcr\noalign{\kern2pt}
$\scriptstyle{#2}$\crcr\noalign{\kern2pt\nointerlineskip}
\kern-2pt$\hfil\displaystyle{#1}\hfil$\crcr}}}\limits}

\def\nn{ \nonumber }
\def\bq{ \begin{equation} }
\def\eq{ \end{equation} }
\def\ben{ \begin{eqnarray} }
\def\en{ \end{eqnarray} }
\def\frac#1#2{{#1\over #2}}
\def\dfrac#1#2{{\displaystyle{#1\over#2}}}

\begin{document}

\title{Commutative Poisson subalgebras for the Sklyanin bracket and deformations of known integrable models}
\author{V. V. Sokolov \\
Landau Institute for Theoretical Physics,\\
Kosygina str. 2, Moscow, 117334, Russia \\
\it\small e-mail: sokolov@landau.ac.ru\\\\
A. V. Tsiganov\\
 Department of Mathematical and Computational Physics,\\
 St.Petersburg University,
  St.Petersburg, 198904, Russia\\
\it\small e-mail: tsiganov@mph.phys.spbu.ru
 }

 \date{}
\maketitle

{\small A hierarchy of commutative Poisson subalgebras for the Sklyanin bracket is proposed. Each of
the subalgebras provides a complete
set of integrals in involution with respect to the Sklyanin bracket. Using different representations of the bracket,
we find some integrable models and a separation of variables for them. The models obtained are deformations
of known integrable systems like the Goryachev-Chaplygin top, the Toda lattice and the Heisenberg model.}

\section{Introduction.}
\setcounter{equation}{0}

Let us consider a $2\times 2$ matrix $T(\lambda)$ which depends polynomially on
the parameter $\lambda$
\bq
 T(\lambda)=\left(\begin{array}{cc}
 A & B \\
 C & D
\end{array}\right),\label{TT}
\eq
whose entries are polynomials of the form
\ben
&A\,(\lambda)= \lambda^{N}+A_{N-1}\,\lambda^{N-1}+\ldots+A_0\,,\qquad
&B\,(\lambda)= B_{N-1}\,\lambda^{N-1} +\ldots+B_0, \label{asymp}\\
 \nn \\
&C\,(\lambda)= C_{N-1}\,\lambda^{N-1} +\ldots+C_0,\qquad
&D\,(\lambda)=D_{N-2}\lambda^{N-2}+\ldots+D_0\,. \nn
\en
The algebra ${\mathfrak A_N}$ of all functions depending on the coefficients
\bq
A_0,\ldots,A_{N-1},~B_0,\ldots,B_{N-1},~C_0,\ldots,
C_{N-1},~D_0,\ldots, D_{N-2}. \label{var}
\eq
of the matrix $T(\lambda)$ is a Poisson algebra with respect to the so called
Sklyanin bracket \cite{skl85}. The dimension of generic simplectic leaves for this bracket is equal
to $2N$.

The main property of the Sklyanin bracket is that the coefficients of the trace of
$T(\lambda)$ generate
an $N$-dimensional commutative subalgebra $\mathfrak A_N^0$ in
 $\mathfrak A_N$. The generators of $\mathfrak A_N^0$ are {\it linear}
polynomials.  In this paper we construct different $N$-dimensional commutative subalgebras
$\mathfrak A_N^M, \,\, M\in \mathbb N,$ in
$\mathfrak A_N$ generated by polynomials of higher degrees. These subalgebras can be regarded as deformations of the
standard trace subalgebra $\mathfrak A_N^0$ . The generators of $\mathfrak A_N^M$ define a set
$S_N^M=\{I_1^M,\cdots,I_N^M\}$ of $N$
integrals in involution. The separation variables for all these sets $S_N^M$ coincide with those for $S_N^0$
whereas the separated curves are different.

Using known representations for the Sklyanin bracket and our families $S_N^M$ of integrals in involution
we can construct
new integrable models. They are deformations of integrable models corresponding to the standard trace integrals.
One of the examples is related to known  representation \cite{skl85,kuzts89} of $\mathfrak A_2$  on the
Poisson algebra $e(3).$

The most interesting models are related to the commutative subalgebras $\mathfrak A_N^1$ possessing a quadratic polynomial in variables
(\ref{var}) which can be regarded as a Hamiltonian. For example, the subalgebra $\mathfrak A_2^1$ gives us the quadratic deformation of
the Goryachev-Chaplygin top found in \cite{soktsig1}. In this paper we construct integrable
deformations of arbitrary degree for
the Goryachev-Chaplygin Hamiltonian.

The well-known Goryachev-Chaplygin case (with the additional gyrostatic term) in rigid body dynamics
is described by the following Hamiltonian
\begin{equation} \label{GOR}
H= J_{1}^2 + J_{2}^2 + 4\, J_{3}^2 + 2 c_1 x_1+2 c_2 x_2+c_3 J_3
,
\end{equation}
where $c_i$ are arbitrary constants. The Lie-Poisson brackets

\begin{equation} \label{e3}
\bigl\{ J_i\,,J_j\,\bigr\}= \varepsilon_{ijk}\,J_k\,, \qquad
\bigl\{ J_i\,,x_j\,\bigr\}= \varepsilon_{ijk}\,x_k\,,\qquad
\bigl\{ x_i\,,x_j\,\bigr\}= 0\,,\qquad i,j,k=1,2,3,
\end{equation}
where $\varepsilon_{ijk}$ is the standard totally skew-symmetric
tensor, defines the corresponding equations of motion. These brackets
possess two Casimir elements
$(x,x)$ and $(x,J)$, where $J=(J_1, J_2, J_3)$, $x=(x_1, x_2, x_3)$ and  $(x,y)$ stands for
the scalar product in ${\mathbb R}^3$.

On the fixed level $(x,J)=0$ of the second Casimir element the Hamiltonian (\ref{GOR}) commutes with an
additional cubic integral of motion. This fact ensures the integrability of the Goryachev-Chaplygin case.

The quadratic deformation of the Goryachev-Chaplygin Hamiltonian given by
\begin{equation} \label{genGOR}
\begin{array}{l}
H=I_1= J_{1}^2 + J_{2}^2 + 4\, J_{3}^2 + 2 c_1 x_1+2 c_2 x_2+c_3 J_3+\\[3mm]
\qquad 4 (a_1 x_1+a_2 x_2)\, J_3-(a_1^2+a_2^2) x_3^2
\end{array}
\end{equation}
also has an additional cubic integral \cite{soktsig1}.
If $c_1=c_2=c_3=0$ then (\ref{genGOR}) gives us a new partially integrable case (i.e. integrable on a
special level $(x,J)=0$ of one of the integrals of motion) for the
Kirchhoff problem of motion of a rigid body in the ideal fluid. A similar deformation of the Kowalewski top
has been considered in
 \cite{sok1,sok2,soktsig1}.

As an application of our general scheme we present below a separation of variables for this model.
The canonical separated variables for the Goryachev-Chaplygin top \cite{skl85} are given by
\bq
q_{1,2}=J_3\pm\sqrt{J_1^2+J_2^2+J_3^2~},\qquad p_{1,2}=\dfrac{1}{2i}\ln
\Bigl(q_{1,2}(ix_1-x_2)-(iJ_1-J_2)x_3 \Bigr)\,.\label{gchsep}
\eq
If we substitute two pairs of variables
 (\ref{gchsep}) into the following separated equation
\bq
\lambda^3+a_0\lambda^2-I_1\lambda+I_0=c_0\mu-\dfrac{b_0\lambda^2\,(x,x)}{\mu}\,,
\qquad \lambda=q_{1,2},\quad\mu=\exp(2ip_{1,2})\,,\label{eqcur1}
\eq
and solve the couple of linear equations obtained with respect to $I_1=H$
 and $I_0$ we immediately
derive the
 Hamilton function
\bq
H_{gch}=J_1^2+J_2^2+4J_3^2+2 a_0 J_3-i (c_0+b_0) x_1+(c_0-b_0) x_2\,.\label{hamgch}
\eq
and additional integrals of motion for (\ref{hamgch}).

Substituting the same variables $(p,q)$ into another separated equation
\bq
\lambda^3+a_0\lambda^2-I_1\lambda+I_0=(c_1\lambda+c_0)\mu
-\dfrac{(b_1\lambda+b_0)\lambda^2\,(x,x)}{\mu}\,,  \label{eqcur2}
\eq
we get an integrable system with the following Hamiltonian
\bq
H=H_{gch}+\Bigl(c_1(iJ_1-J_2)+b_1(iJ_1+J_2)\Bigr)x_3
        -2\Bigl(c_1(ix_1-x_2)+b_1(ix_1+x_2)\Bigr)J_3\,.
        \label{dgch}
\eq
After the canonical transformation
\bq x\to x\,,\quad J\to J+U x\,,\qquad
U=\left(
\begin{array}{ccc}
  0 & 0 & -ic_+ \\
  0 & 0 & -c_- \\
  ic_+ & c_- & 0
\end{array}
\right)\,,\qquad c_\pm=\dfrac{b_1\pm c_1}{2}, \label{ctrin}
\eq
the latter Hamiltonian becomes
\begin{equation} \label{desgch}
\begin{array}{l}
H=J_1^2+J_2^2+4J_3^2+2\Bigl(i(c_1+b_1)x_1-(c_1-b_1)x_2+a_0\Bigr)J_3+c_1b_1x_3^2
\\[4mm]
\qquad -i(b_0+c_0-a_0(c_1+b_1))x_1+(c_0-b_0-a_0(c_1-b_1))x_2\, ,
\end{array}
\end{equation}
which coincides with (\ref{genGOR}).

In Section 4 the simplest deformations for the Toda lattice and the Heisenberg model are given. We do not know
whether physical applications of such deformed models exist.

{\bf Acknowledgments.} The authors are grateful to the Newton Institute (Univ. of Cambridge)
for its hospitality. The
 research was partially supported by RFBR grants 99-01-00294 and 99-01-00698,
INTAS grants 99-1782 and 99-01459, and EPSRC grant GR K99015.

\section{Properties of the Sklyanin bracket}
\setcounter{equation}{0}

The explicit form of the Sklyanin brackets for the coefficients of the matrix (\ref{TT}) can be derived from
the following operator definition
\bq
\{\,\on{T}{1}(\lambda),\,\on{T}{2}(\mu)\}= [r(\lambda-\mu),\,
\on{T}{1}(\lambda)\on{T}{2}(\mu)\,]\,, \label{rrpoi}
\eq
where we use the standard notations $\on{T}{1}(\lambda)=
T(\lambda)\otimes Id\,,~\on{T}{2}(\mu)=Id\otimes T(\mu)$ and
\bq
r(\lambda-\mu)=\dfrac{\eta}{\lambda-\mu}\,\Pi\,,\qquad\mbox{\rm
where}\qquad \Pi=\left(\begin{array}{cccc}
  1 & 0 & 0 & 0 \\
  0 & 0 & 1 & 0 \\
  0 & 1 & 0 & 0 \\
  0 & 0 & 0 & 1
\end{array}\right)\,,\qquad \eta\in {\mathbb C}\,.\label{rr}
\eq

{\bf Example 1.} In the simplest case $N=2$ relation (\ref{rrpoi}) is equivalent to
\ben
&&\{A_1,A_0\}=0, \qquad \{A_1,B_1\}=\eta B_1, \qquad \{A_1,C_1\}=-\eta C_1, \qquad \{A_1,B_0\}=\eta B_0, \nn\\
&&\{A_1,C_0\}=-\eta C_0,\qquad \{A_1,D_0\}=0, \qquad  \{A_0,B_1\}=\eta B_0, \qquad \{A_0,C_1\}=-\eta C_0,\nn\\
&&\{A_0,B_0\}=\eta (B_0 A_1-A_0 B_1), \qquad  \{A_0,C_0\}=\eta (A_0 C_1-C_0 A_1), \label{I1}\\
&&\{A_0,D_0\}=\eta (B_0 C_1-B_1 C_0), \qquad \{B_1,C_1\}=\{B_1,B_0\}=0, \nn\\
&&\{B_1,C_0\}=-\eta D_0, \qquad \{B_1,D_0\}=0, \qquad \{B_0,C_1\}=-\eta D_0, \qquad \{B_0,C_0\}=-\eta D_0 A_1,\nn \\
&&\{B_0,D_0\}=-\eta B_1 D_0, \qquad \{C_1,C_0\}=\{C_1,D_0\}=0, \qquad
\{C_0,D_0\}=\eta C_1 D_0. \nn
\en

It was proven in (\cite{skl85}) that the coefficients
of the determinant
\bq
d(\lambda)=\mathfrak{\rm
det}\,T(\lambda)=A(\lambda)D(\lambda)-B(\lambda)C(\lambda)
\label{Acentre}
\eq
belong to the centre of $\mathfrak A$ or, in other words, they are Casimir elements for
bracket (\ref{rrpoi}). The number of the Casimir functions is $2N-1$ and therefore we have a $4N-1$ dimensional
Poisson manifold with degenerate Poisson structure (\ref{rrpoi}) and $2 N$-dimensional generic symplectic leaves.

To bring the Poisson bracket (\ref{rrpoi}) into canonical form,
a new set of variables
\bq
 \label{newvar}
d_0,\cdots, d_{2 N-2}, \qquad q_1,\cdots q_N, \qquad p_1,\cdots, p_N.
\eq
was proposed in \cite{skl95}.
The variables $d_0,\cdots, d_{2 N-2}$ are the coefficients of (\ref{Acentre}):
$$
d(\lambda)=d_{2N-2}\lambda^{2N-2}+\ldots+d_{0};
$$
and the variables $q_i$ are zeros of the polynomial $A(\lambda)$:\,
$$A(\lambda)=\prod_{j=1}^N (\lambda-q_j);$$
the variables $p_i$ are defined by
$$p_j=\eta\, \ln B(q_j)\,.$$

It follows from (\ref{rrpoi}) that
\[\{A(\lambda),A(\mu)\}=\{B(\lambda),B(\mu)\}=0\]
and
\[
\{A(\lambda),B(\mu)\}=\dfrac{\eta}{\lambda-\mu}\Bigl(A(\lambda)B(\mu)-A(\mu)B(\lambda)\Bigr).
\]
These relations imply $ \{q_j,q_k\}=\{p_j,p_k\}=0$ and
$\{p_j,q_k\}=\delta_{jk}\,$,
respectively.
As usual if we fix values of the Casimir elements $d_j,$ then the canonically conjugated variables $q_i, p_i$
are simplectic coordinates on the corresponding simplectic leaf.

To express the variables (\ref{var}) in terms of (\ref{newvar}), one can use the formulae
\[
\begin{array}{ll}
A(\lambda)=\prod_{j=1}^N (\lambda-q_j),\qquad &D(\lambda)=\dfrac{d(\lambda)+B(\lambda)C(\lambda)}{A(\lambda)}\,,\\
\nn\\
B(\lambda)=\sum_{j=1}^N e^{\eta p_j}\,        \prod_{k\neq j}\left(\dfrac{\lambda-q_k}{q_j-q_k}\right)\,,\qquad
&C(\lambda)=\sum_{j=1}^N d(q_j) e^{-\eta p_j}\,\prod_{k\neq j}\left(\dfrac{\lambda-q_k}{q_j-q_k}\right)\,.
\end{array}
\]

\section{A construction of commutative subalgebras}
\setcounter{equation}{0}

Let us introduce the matrix
\bq
\widetilde{T}(\lambda)=K(\lambda)\,T(\lambda)\,, \label{newlax}
\eq
where $T(\lambda)$
is given by (\ref{TT}),
\bq
 K(\lambda)=\left(\begin{array}{cc}
 {\cal A} & {\cal B} \\
 {\cal C} & 0
\end{array}\right)\,(\lambda),\label{KK}
\eq
whose entries are polynomials of the form
$$
\begin{array}{c}
{\cal A}\,(\lambda)= {\cal A}_{M}\lambda^{M}+{\cal A}_{M-1} \lambda^{M-1}+\cdots+{\cal A}_{0}\,,\\[4mm]
{\cal B}\,(\lambda)= b_{M}\lambda^{M}+b_{M-1} \lambda^{M-1}+\cdots+b_{0}, \qquad
{\cal C}\,(\lambda)= c_{M}\lambda^M+c_{M-1} \lambda^{M-1}+\cdots+c_{0},
\end{array}
$$
and $b_k,c_k\in{\mathbb C}$ are arbitrary parameters. We require that
the trace of the new matrix $\widetilde{T}$ has the form
\bq
\mbox{\rm trace}\, \widetilde{T}(\lambda)=\sum_{i=N}^{N+M} a_{i-N} \lambda^{i}+\sum_{k=0}^{N-1} I_k \lambda^k,
\label{main}
\eq
where $a_i$ are arbitrary fixed constant parameters. It easy to see that both
unknown coefficients ${\cal A}_i$ of the polynomial ${\cal A}$ and unknown functions $I_k$ in (\ref{main}) are
uniquely defined from (\ref{main}). Moreover, they are polynomials in variables (\ref{var}) such that
all the polynomials $I_k$ have the same degree $M+1$ and the degree
of ${\cal A}_i$ equals $M-i$.

{\bf Example 1 (continuation).} If $N=2$ and $M=1$ the functions ${\cal A}_i$ and $I_i$ are given by
\ben
&&{\cal A}_1=a_1, \qquad {\cal A}_0=a_0-a_1 A_{1}-b_1C_{1}-c_1B_{1}\,,\nn\\
\nn\\
&&I_1=a_1(A_0-A_1^2)+b_1(C_0-A_1C_1)+c_1(B_0-A_1B_1)+a_0A_1+b_0C_1+c_0B_1\,,
\label{II1}\\
\nn\\
&&
I_0=(a_0-a_1A_1-b_1C_1-c_1B_1)A_0+c_0B_0+b_0C_0\,.\label{II2}
\en

{\bf Theorem 1.}
Polynomials $I_i$ defined by (\ref{main}) commute with each other with respect to
the bracket (\ref{rrpoi}).

\textbf{Proof.} The explicit form of condition (\ref{main}) is given by
\[
{\cal A}(\lambda)A(\lambda)+{\cal B}(\lambda)C(\lambda)+{\cal C}(\lambda)B(\lambda)
=\sum_{i=N}^{N+M} a_{i-M} \lambda^{k}+\sum_{k=0}^{N-1} I_k \lambda^k.
\]
Let us substitute $\lambda=q_j$ into this identity. It follows from the definition
 $d(\lambda)=AD-BC$ that
\[C(q_j)=-\dfrac{d(q_j)}{B(q_j)}=-d(q_j)\exp(-\eta p_j)\,
.\]
Taking this formula into account we get $N$ linear equations
\bq
{\cal C}(q_j)\,\exp(\eta p_j)-{\cal B}(q_j)\,d(q_j)\,\exp(-\eta
p_j)=\sum_{i=0}^{N-1}{I_i}q_j^i+\sum_{k=N}^{N+M} a_{k-N}\,q_j^k\,, \qquad
j=1,\ldots,N\,.\label{expeq}
\eq
for $N$ unknowns $I_i$. Following \cite{st91}, we can easily show that the functions $I_i$ are in involution
with respect to (\ref{rrpoi}).

Let us rewrite this
 system in the matrix form
\bq
\phi(p,q)=\textbf{S}(q)\,{I}(p,q)+U(q)\,,\label{steq}
\eq
where
\[
\phi_j={\cal C}(q_j)\,\exp(\eta p_j)-{\cal B}(q_j)\,d(q_j)\,\exp(-\eta p_j)\,,\qquad
U_j=\sum_k a_k\,q_j^k
\]
and
\[
{\bf S}=
\left(
\begin{array}{cccc}
1,\quad& q_1,\quad&\cdots&q_1^{N-1}\\
\\
\vdots& \vdots&\ddots&\vdots\\
\\
1,\quad& q_N,\quad&\cdots&q_N^{N-1}\\
\end{array}
\right)
\]
The solution of (\ref{steq}) is given by
\bq
{I}(p,q)=\textbf{W}(q)\,\Bigl(~\phi(p,q)-U(q)~\Bigr)\,,\qquad
\textbf{W}(q)=\textbf{S}^{-1}(q). \label{gint}
\eq

The matrix $\textbf{S}$ belongs to the class of so-called St\"ackel matrices \cite{st91}
for which the entries $S_{ij}$ of the $i$-th row are functions depending on $q_i$ only.
Differentiating the identity
\[\sum_{i}W_{mi}S_{il}=\delta_{ml}\,,\]
with respect to $q_j$ and using the main property of the St\"ackel matrices,
we have
\[\sum_{i}\dfrac{\partial W_{mi}}{\partial q_j} S_{il}=-W_{mj}\dfrac{\partial S_{jl}}{\partial q_j} \,.\]
Substituting $k$ for $m$ and eliminating the right hand side $\partial S_{jl}/\partial q_j$ from
the two obtained equations, we find
\[\sum_i \left(W_{jk}\dfrac{\partial W_{im}}{\partial q_j} -W_{jm}\dfrac{\partial
W_{ik}}{\partial q_j}\right) S_{il}=0\,.\]
Since the determinant of the matrix ${\bf S}$ does not vanish
\begin{equation}\label{WW}
W_{jk}\dfrac{\partial W_{im}}{\partial q_j} -W_{jm}\dfrac{\partial
W_{ik}}{\partial q_j}=0
\end{equation}
for all $j,k,i,m$.

Calculating the Poisson brackets between $I_k$ and $I_m,$ we obtain
\bq
\{{I}_k,{I}_m\}=\sum_{i,j}
\left(W_{jk}\dfrac{\partial W_{im}}{\partial q_j} -W_{jm}\dfrac{\partial
W_{ik}}{\partial q_j}\right)\dfrac{\partial \phi_j}{\partial
p_j}\Bigl(\phi_j-U_j(q_j)\Bigr)\,.
\nn
\eq
It follows from (\ref{WW}) that
$\{{I}_k,{I}_m\}=0\,.$
The proof is complete.

{\bf Collorary 1.} Variables $p_i,q_i$ are the separated variables for integrals of motion $I_j$
satisfying the separated equations (\ref{expeq})-(\ref{steq}) whereas variables
$$
\lambda_i=q_i,\qquad \mu_i={\cal C}(q_i)\, \exp{(\eta \,p_i)}
$$
satisfy the characteristic equation $Det(\widetilde{T}(\lambda)-\mu)=0$. This means that these variables lie on
the corresponding equivalent algebraic curves.

{\bf Remark 1.} Theorem 1 for $M=0$ is a well-known fact. In the case
$N=2, M=1$ the matrix $\widetilde{T}$ with
\bq \label{KK1}
K(\lambda)= \left(
\begin{array}{cc}
  \lambda+{\cal A}_{0}&
 b_{0}\\
c_{0}& 0
\end{array}\right), \qquad {\cal A}_{0}=a_0-A_1,
\eq
has been proposed in \cite{skl85} in a non-factorized form. In the paper \cite{ts97} this matrix
was factorized and generalized to the case of arbitrary $N$. Notice, that our matrix $K$ for $M=1$
has a more general form
$$
K(\lambda)= \left(
\begin{array}{cc}
  \lambda+{\cal A}_{0}&
b_1 \lambda+ b_{0}\\c_1 \lambda+
c_{0}& 0
\end{array}\right), \qquad {\cal A}_{0}=a_0-A_1-b_1 C_1-c_1 B_1,
$$
than (\ref{KK1}).
Parameters $b_1$ and $c_1$ are absolutely essential in constructing new quadratic integrable Hamiltonians.

{\bf Remark 2.} To generalize the condition (\ref{main}) we can assume that
\bq
\mbox{\rm trace}\,\widetilde{T}(\lambda)=\sum_{k\in\widehat{\varrho}}
a_k\lambda^{k}+\sum_{i\in
\varrho} {I}_i\lambda^{i}, \qquad a_k\in{\mathbb C}\,. \label{cond2}
\eq
Here $\varrho=\{i_1,\ldots, i_N\}$ is an arbitrary subset in the set
of numbers $\{0,1,\ldots,N+M\}$ and $\widehat{\varrho}$ is the
corresponding complement such that $\varrho\cup
\widehat{\varrho}=\{0,1,\ldots,N+M\}$.
 But in this case ${I}_i$ are rational
functions of variables (\ref{var}).

{\bf Example 1 (continuation).} If $N=2$ and $M=1$ and
\[\mbox{\rm trace}\,\widetilde{T}(\lambda)=a_1\lambda^3+I_1\lambda^2+a_0\lambda+I_0\,,\]
these functions are
\ben
&&{\cal A}_1=a_1,\qquad {\cal A}_0=\dfrac1{A_1} \Bigl(a_0-a_1A_0-c_0B_1-b_0C_1-b_1C_0-c_1B_0\Bigr)\,,\label{ratgch}\\
\nn\\
&&I_1= \dfrac{1}{A_1}\Bigl(a_0-a_1(A_0-A_1^2)-b_1(C_0-C_1A_1)-c_1(B_0-B_1A_1)-c_0B_1-b_0C_1\Bigr),\nn\\
\nn\\
&&I_0=\dfrac{1}{A_1}\Bigl(A_0(a_0-c_1B_0-a_1A_0-b_1C_0)+b_0(C_0A_1-A_0C_1)+c_0(B_0A_1-A_0B_1)\Bigr)\,.\nn
\en

{\bf Remark 3.} Theorem 1 can be proven in the same way for matrices (\ref{TT}) of slightly different structure.
For example, we can assume that the entry $D(\lambda)$ of the matrix $T(\lambda)$ has the form
\[D(\lambda)=\lambda^N+D_{N-1}\lambda^{N-1}+D_{N-2}\lambda^{N-2}+\cdots+D_0\,.\]
In this case we have one additional variable $D_{N-1}$ and one additional Casimir function
$d_{2N-1}=A_{N-1}+D_{N-1}$.

\section{Polynomial deformations of known integrable models}
\setcounter{equation}{0}

If we identify the Sklyanin bracket (\ref{rrpoi}) with a fixed Poisson bracket
on some phase space ${\cal M}$, then (\ref{rrpoi}) can be regarded as
an \textit{equation} for matrix $T(\lambda)$ of the form (\ref{TT}) whose entries
(\ref{asymp}) are polynomials in $\lambda$ with coefficients being
functions on
 ${\cal M}$. Such a matrix $T(\lambda)$ defines a representation of (\ref{rrpoi})
on ${\cal M}$.

Using known representations and Theorem 1, we can produce hierarchies of deformations for several
integrable models such as the Goryachev-Chaplygin top, the Toda lattice, and the Heisenberg magnetic.
According to Collorary 1 integrals of motion for all members of such a hierarchy
are separable in the same canonical variables $p_i,q_i$.

\subsection{Deformed Goryachev-Chaplygin top}
Let us consider the Lie algebra
$e^*(3)$ with a natural Lie-Poisson bracket
(\ref{e3}).
The phase space of the Goryachev -Chaplygin top $\cal M$ is a union of
special non-generic coadjoint orbits (symplectic
leaves)
 of $E(3)$ in $e^*(3)$ specified by the fixed value  $I_2=0$ of the second Casimir operator.

It was observed in \cite{kuzts89} that the functions
$$
\begin{array}{c}
\displaystyle A_1=-2 J_3, \qquad A_0=-J_1^2-J_2^2-\frac{\varepsilon}{x_3^2}, \qquad B_1=x_2+i x_1,
\qquad B_0=-(J_2+i J_1) x_3, \\[4mm]
C_1=-x_2+i x_1, \qquad C_0=(J_2-i J_1) x_3, \qquad D_0=x_3^2
\end{array}
$$
satisfy (\ref{I1}) if the Poisson bracket is given by (\ref{e3}), $I_2=0$ and $\eta=-2 i$. Formulae (\ref{II1}) and
(\ref{II2}) give us two integrals $I_1$ and $I_0$ such that $\{I_1,\, I_0\}=0$. It easy to verify that
(up to constant factors)
\begin{equation}
\begin{array}{l}
\displaystyle H_1=I_1=a_1 \Big(J_1^2+J_2^2+4 J_3^2+\frac{\varepsilon}{x_3^2} \Big)+2 a_0 J_3+b_0(x_2-i x_1)-c_0 (x_2+i x_1)
\\[4mm]
\qquad +b_1 (2 J_3 x_2-J_2 x_3-2i J_3 x_1+i J_1 x_3)-c_1(2 J_3 x_2-J_2 x_3+2i J_3 x_1-i J_1 x_3)
\end{array}
\end{equation}\label{hamH1}
which is equivalent to (\ref{genGOR}).
The explicit form of the integral $I_0$ is given by
\begin{equation}
\begin{array}{l}
I_0=2 a_1 (J_1^2+J_2^2) J_3+a_0 (J_1^2+J_2^2)+b_0 (-J_2+i J_1)x_3+c_0 (J_2+i J_1)x_3\\[3mm]
\qquad +b_1(J_1^2+J_2^2)(x_2-ix_1)-c_1(J_1^2+J_2^2)(x_2+i x_1)\\[4mm]
\displaystyle
\qquad +\varepsilon \frac{2 a_1 J_3+a_0+b_1 (x_2-ix_1)-c_1(x_2+i x_1)}{x_3^2}.
\end{array}
\end{equation}

As examples we also present explicitly one  polynomial deformation of higher degree and one rational deformation.
Thus if $M=2$ the deformation reads as follows:
\ben
H_2&=&H_1+4a_2J_3\Bigl((J_1^2+J_2^2+2J_3^2)+\varepsilon x_3^{-2}\Bigr)\nn\\
\nn\\
&+&b_2\Bigl(
2iJ_3x_3(J_1+iJ_2)+(J_1^2+J_2^2+4J_3^2+\varepsilon x_3^{-2})(x_2-ix_1)
\Bigr)\nn\\
\nn\\
&-&c_2\Bigl(
(2iJ_3x_3(J_1-iJ_2)+(J_1^2+J_2^2+4J_3^2+\varepsilon x_3^{-2})(x_2+ix_1)
\Bigr),\nn
\en
where $H_1$ is defined by (\ref{hamH1}).
In the case $M=1$ the simplest rational deformation
\[
\tilde{H}=\dfrac{H_1-a_0(2J_3-1)}{2J_3}\,
\]
corresponds to (\ref{ratgch}).

\subsection{Deformed Toda lattices}
The Sklyanin bracket may be identified \cite{skl85t} with the standard bracket in ${\cal M}={\mathbb R}^{2N}$
with the help of the following ansatz for the matrix $T(\lambda)$:
\bq
T(\lambda)=L_N\,L_{N-1}\cdots L_1\,,\qquad L_i(\lambda)=\left(\begin{array}{cc}\lambda-p_i&e^{q_i}\\ -e^{-q_i}&0
\end{array}\right)\, \qquad q_{i+N}=q_i.\label{todalax}
\eq
Here $p_i,q_i$ are canonical variables in ${\mathbb R}^{2N}$.

The matrix $T(\lambda)$ obeys the Sklyanin bracket (\ref{rrpoi})
with $\eta=1$ and describes the periodical Toda lattice \cite{skl85t}.
The trace of the matrix $T(\lambda)$ produces the integrals of motion. The simplest of them are
\begin{equation}
P= -\sum_{i=1}^N p_i\,,\qquad
H= -\sum_{i>j}^N
p_ip_j -\sum_{i=1}^N e^{q_{i+1}-q_{i}}\,.\label{anh}
\end{equation}

The matrix (\ref{todalax}) has the necessary polynomial structure
(\ref{asymp}) and we may apply Theorem 1 in order to produce deformations
of the periodic Toda lattice, polynomial in momenta.  For instance, if $M=1$ then
the integral of the lowest degree in trace $\widetilde{T}(\lambda)$
is the following quadratic integral
\bq
I_{N-1}=-a_0\sum_{i=1}^N p_i-a_1\Bigl(\sum_{i>j}^N
p_ip_j +\sum_{i=1}^N e^{q_{i+1}-q_{i}}\Bigr)
+e^{q_1}(c_0+c_1p_1)-e^{-q_N}(b_0+b_1p_N)\,.
\label{dtodah}
\eq

\subsection{Deformed spin chain}
Let the phase space $\cal M$ be a direct sum of the algebras $sl^*(2)$ with the following Lie-Poisson bracket
\bq
\{s_3^i,s_\pm^i\}=\pm s_\pm^i\,,\qquad \{s_+^i,s_-^i\}=2s_3^i
\label{gs2}
\eq
in the each algebra. The Sklyanin bracket with $\eta=1$ is identified with (\ref{gs2})
by
\[
T(\lambda)=L_N\,L_{N-1}\cdots L_1\,,\qquad L_i(\lambda)=\left(\begin{array}{cc}\lambda
+s_3^i&s_-^i\\ s_+^i&\lambda-s_3^i
\end{array}\right)\,.
\]
It is easy to check that such a product has the polynomial structure described in Remark 3 under the restriction
that the Casimir fumction $d_{2n-1}=A_{N-1}+D_{N+1}$ is equal to zero. Adapting our construction to this case
we can construct polynomial deformations for the spin chain. In the simplest case $M=1$ we obtain the following
integrable quadratic Hamiltonian
\ben
I_{N-1}=H&=&a_0S_{3}+b_0S_{+}+c_0S_{-}
+a_1\Bigl(\sum_{j>i}^N (s_+^i s_-^j+s_3^is_3^j)-S_3^2\Bigr)\,\nn\\
\nn\\
&-&2b_1\Bigl(S_3S_+-\sum_{j>i}^N s_3^is_+^j+\dfrac12 \sum_{i=1}^N s_3^is_+^i\Bigr)\nn\\
\nn\\
&-&2c_1\Bigl(S_3S_--\sum_{j<i}^N s_3^is_-^j+\dfrac12 \sum_{i=1}^N s_3^is_-^i\Bigr),\nn
\en
where
$$
S_3=A_{N-1}=\sum s_3^i, \qquad S_-=B_{N-1}=\sum s_-^i,\qquad S_+=C_{N-1}=\sum s_+^i.
$$
If $b_1=c_1=0$ then $H$ coincides with the general quadratic integral for the standard spin chain.

It is well-known that the Hamiltonians for the Toda lattice and the spin chain derived from the factorised form of
operator $T(\lambda)$ have to be transformed to bring them to a local form suitable for applications. We do not
know such transformations for the deformed models.


\begin{thebibliography}{10}

\bibitem{kuzts89}
V.B. Kuznetsov and A.V. Tsiganov,
A special case of Neumann's system and the
Kowalewski-Chaplygin-Goryachev top,
\newblock {\em J.Phys.}, v.22, p.L73, 1989.

\bibitem{skl85}
E.K. Sklyanin,
The Goryachev-Chaplygin top and the method of the inverse scattering problem.
Differential geometry, Lie groups and mechanics, VI.\newblock{\em Zap. Nauchn. Sem. LOMI.}, v.133, p.236, 1984.

\bibitem{skl85t}
E.K. Sklyanin,  The quantum Toda chain.,
 Nonlinear equations in classical and quantum field theory
(Meudon/Paris, 1983/1984), 196--233,
\newblock{\em Lecture Notes in Phys.}, v.226, Springer, Berlin, 1985.

\bibitem{skl95}
E.K. Sklyanin,
Separation of variables---new trends. Quantum field theory, integrable models
and beyond (Kyoto, 1994).
\newblock{\em Prog. Theor. Phys. (Suppl)}, v.118, p.35, 1995.


\bibitem{sok1}
V.V. Sokolov,
\newblock{A new integrable case for the
Kirchhoff equation},
\newblock{\em Teor.Math.Phys.}, v.128(2), p.31, 2001.

\bibitem{sok2}
V.V. Sokolov,
\newblock{A generalized Kowalevski Hamiltonian and new integrable
cases on
 $e(3)$ and $so(4)$},
\newblock{ Preprint \em nlin.SI/0110022},  2001.

\bibitem{soktsig1}
V.V. Sokolov,
 A.V. Tsiganov,
On the Lax pairs for the generalized Kowalewski and Goryachev-Chaplygin tops,
\newblock{\em Teor.Math.Phys.}, v.(), p., 2002.

\bibitem{st91}
P. St\"{a}ckel, \"{U}ber die integration der Hamilton-Jacobischen differentialgeichung
mittels separation der variabeln, Habilitationschrift, Halle, 1891.
\\
L.A. Pars, An elementary proof of the St\"{a}kel theorem,
\newblock{\em American Math. Monthly}, v.56, p.394, 1949.

\bibitem{ts97}
A.V. Tsiganov,
The Kowalewski top, a new Lax representation.
\newblock {\em J.\-Math.\-Phys.}, v.38, p.196, 1997.


\end{thebibliography}
\end{document}